# Nanoscale plasticity and neuromorphic dynamics in silicon suboxide RRAM

*Mark Buckwell\*, Wing H. Ng, Daniel J. Mannion, Stephen Hudziak, Adnan Mehonic, and Anthony J. Kenyon*

Dr. M. Buckwell, Dr. W. H. Ng, Mr. Daniel J. Mannion, Dr. Stephen Hudziak, Dr. A. Mehonic, Prof. A. J. Kenyon
Department of Electronic and Electrical Engineering, University College London, London, WC1E 7JE, United Kingdom
E-mail: markbuckwell@ucl.ac.uk

Resistive random-access memories, also known as memristors, whose resistance can be modulated by the electrically driven formation and disruption of conductive filaments within an insulator, are promising candidates for neuromorphic applications due to their scalability, low-power operation and diverse functional behaviours. However, understanding the dynamics of individual filaments, and the surrounding material, is challenging, owing to the typically very large cross-sectional areas of test devices relative to the nanometre scale of individual filaments. In the present work, conductive atomic force microscopy is used to study the evolution of conductivity at the nanoscale in a fully CMOS-compatible silicon suboxide thin film. Distinct filamentary plasticity and background conductivity enhancement are reported, suggesting that device behaviour might be best described by composite core (filament) and shell (background conductivity) dynamics. Furthermore, constant current measurements demonstrate an interplay between filament formation and rupture, resulting in current-controlled voltage spiking in nanoscale regions, with an estimated optimal energy consumption of 25 attojoules per spike. This is very promising for extremely low-power neuromorphic computation and suggests that the dynamic behaviour observed in larger devices should persist and improve as dimensions are scaled down.

# 1. Introduction

Binary oxides are promising materials for low power, high packing density resistance switching devices (sometimes referred to as memristors, or resistive random access memory, RRAM), with the potential for both non-volatile memory applications, and in-memory and neuromorphic computation.[1–3] RRAM offers the exciting prospect of implementing computational functionality at the device level, drastically reducing power and real-estate requirements, rather than requiring complex CMOS integration and control. To date, RRAM devices and arrays have been used to perform vector matrix multiplication, to implement training and image recognition in neural networks, and in more bespoke applications such as edge detection,[4–7] although this list is non-exhaustive.

The functionality of these oxides is dependent upon their behaviour under electrical stress, with conductivity changes generally thought to result from changes in oxidation state[8] which, when localized, lead to the formation of nanoscale conductive channels known as filaments.[9,10] Understanding filamentary dynamics remains challenging, as it is difficult to precisely probe the electrical behaviour of individual filaments.[11,12] Ultimately, to develop RRAM technologies for high-density memory and computation, it is necessary to determine whether functionalities are maintained as device dimensions are minimized.[13]

Scanning probe techniques have proven useful in studying the conduction mechanisms and breakdown modes of gate oxides at the nanoscale,[14–17] as well as RRAM materials and devices.[18–23] In this work, we use conductive atomic force microscopy (CAFM) as a nanoscale electrical probe for spatial and temporal measurements of the evolution and dynamics of conductivity changes in silicon suboxide, $SiO_x$. Our devices are fully CMOS-compatible, and have excellent non-volatile memory properties, such as low switching voltages, high endurance and long retention, as well as desirable neuromorphic dynamics such as thresholding, spiking and plasticity.[24,25] Together, these functionalities can produce a diverse set of behaviours that emulate the connectivity and signalling of the human nervous system. Using CAFM, we show that individual filaments also demonstrate these behaviours, suggesting that the functionality of our devices should persist as they are scaled down. We also find that there is a dynamic background conductivity, which might play a crucial role in device performance. Finally, we present a study of filamentary voltage spiking indicating that an idealized device array might have an ultra-low power consumption of less than 1 $\mu Wcm^{-2}$, significantly lower than that of the human brain.

It should be mentioned that the switching mechanism for our devices,[8,26–28] and for binary oxide RRAM devices in general,[29,30] has been discussed extensively, including the correlation between structural and electrical changes, so in this work we will not discuss it in detail. For clarity, and for further details, the reader is directed to the extensive literature on the subject. Rather, we briefly mention here that our devices to behave according to a model in which an applied electric field produces defects in the active layer, likely oxygen vacancies. With sufficient accumulation, these defects form a conductive filament between the top and bottom electrode. This filament may then be disrupted by Joule heating under current injection.[8,31] Under electrical stress there is an intrinsic interplay between these processes, with associated deformations occurring in the active layer and electrodes as a result of oxygen migration.

## 2. Experimental section

A 65 nm molybdenum bottom electrode layer was sputter-deposited on a p-type silicon wafer capped with 1 μm thermally grown $SiO_2$. An 11 nm $SiO_x$ layer was then deposited by sputtering a silicon target in an oxygen atmosphere, giving x ≈ 1.6. CAFM measurements were carried out with a Bruker Icon microscope under ambient conditions using platinum-coated (Spark, NuNano and SCM-PIC, Bruker) and full-platinum (RMN-Pt, Rocky Mountain Nanotechnology) probes. The nominal spring constant and deflection sensitivity were used to estimate that the applied force was no more than 100 nN.[32] Anodic oxidation was avoided as electrons were injected from the bottom molybdenum electrode (i.e. the stage was negatively biased with respect to the probe) and all applied voltages had a magnitude of no more than 10 V.[14,33–35] Data were processed and analysed using Nanoscope Analysis v1.5, as well as Matlab (for spike counting and width classification using the function findpeaks, and fitting of spikes per second and energy per second using nonlinear and linear least squares regression, respectively). All reported voltage and current values are absolute. Note that the current detector saturation is not a limit; the real current can exceed saturation, which is around 12.3 nA at an instrumental current sensitivity (i.e. amplifier gain) of 1 $nAV^{-1}$ or 420 nA at a sensitivity of 100 $nAV^{-1}$.

In the body of this manuscript we focus on the electrical characterization of material functionality, and do not present the topographical data associated with the current data from our CAFM measurements. This is because a full description of the structural changes occurring, and their relation to the associated switching mechanism, has already been discussed extensively, as noted in our introduction. Nevertheless, in the interests of completeness, we have made some comments on topographical changes in the text, and included topography images, along with more detailed discussion, in the Supporting Information.

## 3. Results and discussion

### 3.1 Spatially resolved plasticity

Synaptic plasticity describes the enhancement and suppression of a connection between neurons or, in the case of an RRAM device, a pair of electrodes.[36,37] Potentiation (enhancement) and depression (suppression) of the connection strength, or synaptic weight, is essential in biology for behavioural learning and, in electronic devices, is typically achieved by applying voltage pulses, ideally with fixed characteristics (polarity, magnitude, shape, duration).[3] We have previously demonstrated plasticity in $SiO_x$ devices when applying such pulses using a semiconductor parameter analyser.[25,38]

Our $SiO_x$ is amorphous, although we have observed that it has an important microstructure; density variations appear to produce defect-rich columnar features and interface roughness that facilitate filament growth.[8,39,40] Indeed, the behaviour of many RRAM devices is dependent on the formation of localized filaments in the active layer, rather than a bulk effect.[19–21,41,42] It is therefore of great interest to achieve a spatial characterization of plasticity, in order to determine the nanoscale dynamics that enable such crucial functionality. With CAFM, this may be done by observing the evolution of the current in a scanned region over time, between successive scans, under the application of a constant electrical stress.

To this end, we used CAFM to scan a 500 nm × 500 nm area repeatedly with a constant voltage stress while mapping current, as shown schematically in **Figure 1**a. We applied 7.7 V, which was sufficient to produce contrast in the current images and induce persistent conductivity changes. Lower voltages did not produce these effects; we suppose this is because the probe is only briefly in contact with any point on the surface. In effect, we applied lines of voltage pulses, or 'pseudopulses', of a single magnitude and polarity to our sample. Figure 1b to e show the evolution of current in the scanned area over successive scans (the full set of scan topographies and currents is shown in Supplementary Figure S1). We note two significant features; the entire area becomes more conductive, and two highly conductive spots appear, corresponding to the tops of conductive filaments. It is worth noting that these features appear rather wide, up to around 100 nm, although their size is likely overestimated due to the size of the scanning probe and the saturated colour bar.

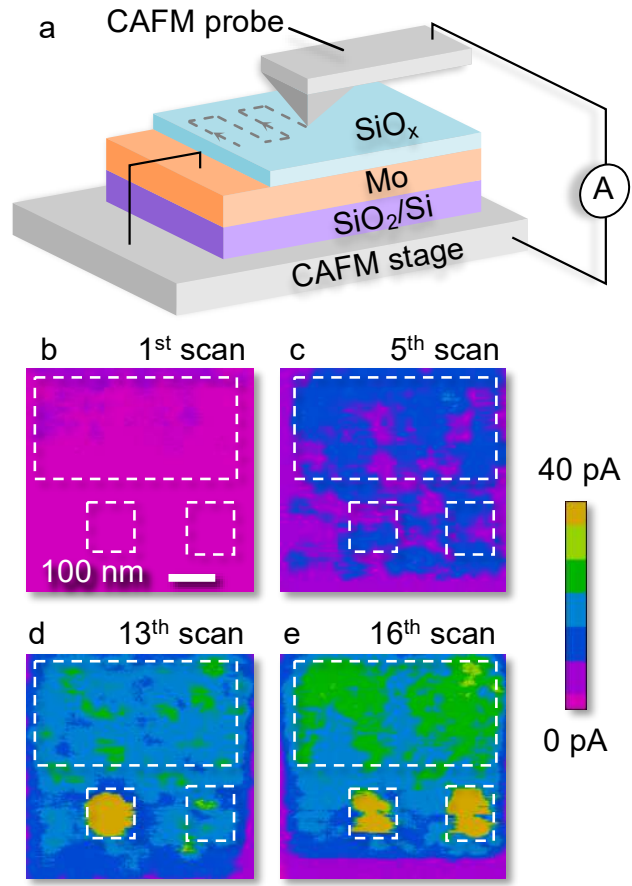

**Figure 1.** Contrasting background and filament conductivity dynamics at 7.7 V in our $SiO_x$. a) Schematic of CAFM instrumentation. b) to e) Evolution of the current in the 500 nm × 500 nm scanned area. The whole area becomes more conductive with successive scans, and two filaments appear. Note the colour bar saturates at 40 pA, though the current in the filaments reaches saturation, at 12.3 nA. The dashed white boxes indicate the background and filament regions sampled for Figure 2.

Given an estimated probe diameter of 40 nm and accounting for edge convolution, we estimate the true diameter of these features to be around 14 nm (a description of the estimation method is presented in Supplementary Figure S2). We also observe that the topography of the entire scan area becomes smoother over the course of the measurements (Supplementary Figure S1). We believe this to be caused by deformation of the $SiO_x$ and/or the molybdenum, rather than by blunting of the CAFM probe, which retained its imaging quality for subsequent scans (Supplementary Figure S3). Additionally, the presence of localized features which change gradually in contrast to each other and the surrounding area suggests that we are imaging a real sample behaviour, rather than an instrumental artefact.

RRAM behaviour is predominantly attributed to filament dynamics,[1] in some cases in parallel with a static background resistance or capacitance,[27] as we have also previously observed in $SiO_x$.[43] However, our observations in Figure 1 demonstrate that the background may also be dynamic, as the whole scan area was modified by electrical stress. $NbO_x$ RRAM devices have been reported to conduct via a core (filament) in parallel with a shell (background).[44,45] The combination of two parallel channels can give rise to complex oscillatory dynamics – important for neuromorphic applications. We suggest that such a core/shell model might be valid for many RRAM materials. An important consideration that emerges from this previous work is that dynamic device performance and functionality might require both a core and a shell, although in smaller devices, close to the size of an individual filament, the background component may disappear. We add the caveat that, here, the CAFM probe is a mobile electrode and does not address the whole scan area simultaneously.

**Figure 2**a shows the evolution of the background current compared to that through each of the filaments shown in Figure 1b to e. Up to scan 8, each region demonstrates a similar increase in mean current (i.e. the average of the pixels enclosed by the dashed boxes in Figure 1b to e). This is schematically illustrated in Figure 2b, wherein the electric field produces defects across the whole scan area, increasing the conductivity of the active layer. This is consistent with our previous report that electrically biasing silicon oxide results in large scale changes in the distribution of oxygen.[8] After scan 9, the maximum current in the left-hand filament increases rapidly, saturating the detector in scan 11. Notably, the mean current lags the maximum by 2 to 3 scans. The high maximum current suggests the full formation of a filament through the $SiO_x$, as shown in Figure 2c.

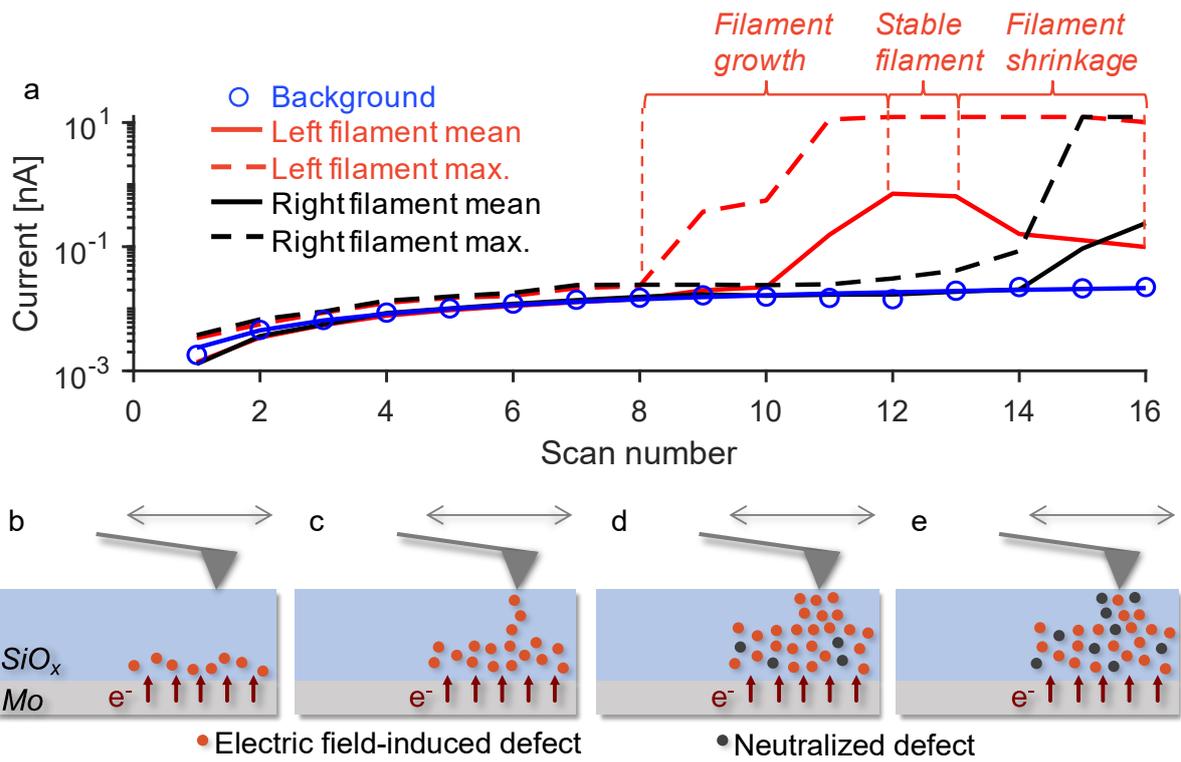

**Figure 2.** a) Comparison of currents through the background matrix, and left- and right-hand filaments at a constant bias of 7.7 V, corresponding to the regions denoted by the dashed boxes in Figure 1b to e. b) to e) Schematic representations of the growth and rupture of the left-hand filament, as defects are induced by the electric field and neutralized by Joule heating.

The slower increase in mean current suggests lateral filament growth, as the region showing high current expands (as illustrated in Figure 2d and shown experimentally in Supplementary Figure S1). We can describe these processes as a gradual potentiation of a connection between the CAFM probe and the molybdenum bottom electrode. Scans 12 and 13 demonstrate stable current, corresponding to a stable filament. However, from scan 14 onwards, the mean current decreases, though the maximum remains at saturation. This suggests that the filament may be narrowing, as shown in Figure 2e, corresponding to depression of the connection between the probe and bottom electrode, a regime in which Joule heating induces sufficient defect neutralization, through re-oxidation, disrupting the filament but not fully breaking it. We also note the onset of potentiation at scan 12 for the right-hand filament. We expect that the difference in the behaviour between the two filaments (i.e. the number of scans required to potentiate the sample at each locations) results from the intrinsic stochasticity of filament formation, arising from the amorphous nature of the $SiO_x$ (local roughness and density variations).

Following the disruption of the left-hand filament in Figure 1 and Figure 2, we scanned a larger area at a reduced stress of 5.5 V to try to pause the evolution of the right-hand filament. This voltage was chosen as it was sufficient to produce contrast in the current imaging between the surrounding, pristine sample area and the stressed region, without inducing any further changes (Supplementary Figure S4 demonstrates that voltages between 1V and 5 V did not produce any measurable current). **Figure 3**a demonstrates the current map after a single reading scan at a reduced stress of 5.5 V. Both filaments are still intact, although their positions are slightly different to Figure 1e. It is possible that we are only observing the most conductive part of each filament because the reading voltage, and thus the current, is much lower. The position and shape of the left-hand filament, in particular, is quite different. This might be a charging artefact or the result of surface contamination, or perhaps the appearance of a new filament, although this is unlikely as the reduced voltage was insufficient to produce any other change in conductivity. Topographical images show that the entire scan area, including that scanned only at 5.5 V, appears smoother than the pristine $SiO_x$ (Supplementary Figure S4 and Supplementary Figure S5). We expect this to result from deformation of the $SiO_x$ and/or molybdenum, rather than probe blunting, as we have previously discussed (i.e. that the imaging quality of this probe was maintained, as shown in Supplementary Figure S3). Interestingly, this may suggest that some structural change occurs at voltages below 7.7 V, below that required to observe pronounced changes in electrical behaviour. Alternatively, those changes occurring at 7.7 V may have a de-localized distribution within the active layer or bottom electrode, affecting an area larger than the addressed region.

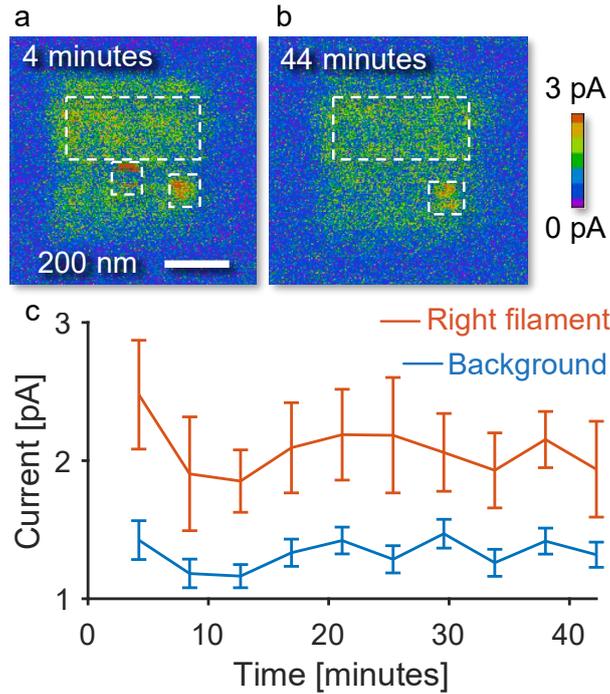

**Figure 3.** Persistence of the filament and background conductivity. a) Current map of the area stressed in Figure 1, following a single reading scan at 5.5 V. The central area (green) is more conductive than the surroundings (blue), with higher currents at the location of each filament (red), the locations of which are indicated by the smaller white boxes. b) Current map 40 minutes after a. The current in both the stressed region and the right-hand filament are unchanged, but the left-hand filament has disappeared. c) The mean current in the stressed region and the right-hand filament, as indicated by the dashed white boxes in a and b, over 40 minutes of reading. There is no significant change to either the filament or the stressed region.

The previously stressed area in Figure 3a is more conductive that its surroundings. Figure 3b shows the area following an additional 40 minutes of reading scans (10 consecutive scans in total). The background current and right-hand filament appear consistent with Figure 3a, but the left-hand filament has disappeared; this occurred after 2 scans. Figure 3c shows the mean current in the stressed region and the right-hand filament over the course of the reading measurements, neither of which changes significantly. Thus, we have been able to potentiate and depress a filament through the use of CAFM pseudopulses, and to stabilize a potentiated filament by reducing the applied voltage.

The consistency of the current in the stressed area during the reading measurements suggests that capacitive charging has not occurred, as we would expect the current to decrease over time if this was the case. Furthermore, such areas do not exhibit evidence of discharging when the CAFM probe is held in a fixed position, applying a constant voltage stress in either polarity. Therefore, the increase in conductivity appears to be a real, persistent effect. This further supports a core/shell model with a significant parallel background component.

**3.2 Neuromorphic filamentary dynamics**

To more directly study the temporal dynamics of filaments in our $SiO_x$, we used CAFM to apply constant voltage stresses at fixed locations. **Figure 4**a demonstrates the typical time dependence of the current under a constant stress of 4 V. Initially, there is no measurable deviation in current from the noise floor (period i). The current then begins measurably increasing, initially quite slowly (period ii), then more rapidly (period iii); as a filament begins to form, the effective dielectric thickness will decrease and so the electric field will increase, causing positive feedback in the filament formation. Finally, the current jumps beyond 1 nA, then fluctuates across nearly two orders of magnitude and eventually saturates the detector (period iv). In this final period, we observe the competition between filament formation and rupture. We note that over the long duration of this measurement, some lateral drift occurred, which might lead to an overestimate of the time taken for the current to change (as shown in Supplementary Figure S6). It is also interesting that the topography of the local area appears consistent before and after the application of the voltage stress, although the addressed location shows a slight deformation. This leads us to conclude that the probe was not significantly affected by the measurement (i.e. no significant blunting or platinum deposition occurred, as imaging quality is maintained), although the $SiO_x$ or molybdenum might have deformed slightly.

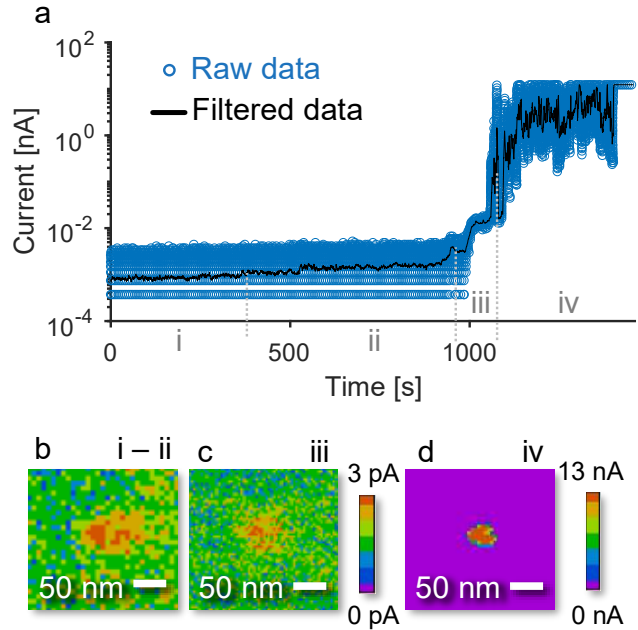

**Figure 4.** Time dependence of the current under constant voltage stress in SiO$_x$. a) With 4 V applied, the current increase accelerates over periods i to iii, eventually reaching instability and detector saturation in period iv. The raw data points are very dense, so data processed with a 1 second wide Savitzky-Golay filter are shown for clarity. b) to d) Evolution of the filament location during each period, taken from a different measurement to that shown in a.

We imaged the top of a filament periodically during voltage stressing to assess its appearance following the application of increasing current, corresponding to the periods shown in Figure 4a (i.e. a different filament to that shown in Figure 4a). We initially applied 4.5 V to produce a few pA (equivalent to period i – ii), then 6 V to produce around 10 pA (equivalent to period iii), and 6 V again to produce current fluctuations between 100 pA to 10 nA (equivalent to period iv). Figure 4b to d shows current maps at equivalent periods to those shown in Figure 4a. Following periods i and ii, we observe a spot of a few pA above background when scanning at 7.8 V. We note that this voltage might affect the conductivity of the SiO$_x$, although we were unable to resolve the spot at lower bias. However, from the trend in Figure 2a, we would not expect the effect of imaging at 7.8 V to be significant compared to the effect at the contact location when the probe is held in place for a longer duration at lower bias. The appearance of the spot does not change significantly following period iii; the conductivity increase appears not to persist once the voltage is removed. Finally, in period iv, the spot becomes much more pronounced and we measure a current at the saturation level of 12.3 nA. This highly conductive spot persisted at the measured location, such that we were able to perform current-voltage sweeps to ± 50 mV, giving us Ohmic behaviour below saturation, and a resistance of around 2.3 kΩ, although we note that the contact resistance between the probe and sample is unknown. Thus, although the current is unstable in period iv, this region corresponds to the presence of a filament across the $SiO_x$. In fact, it is this rich, dynamic interplay between filament formation and rupture that gives rise to valuable neuromorphic functionalities such as spiking.[24] Corresponding topography images for this measurement demonstrate the appearance of a transient distortion on the sample surface (Supplementary Figure S6). We expect this to be either a charging artefact or a deformation of the $SiO_x$ and/or molybdenum that subsequently relaxes. Otherwise, the local area retains its appearance, so we do not expect significant probe blunting to have occurred.

We have previously observed neuromorphic spiking in larger devices.[27] In both biology and electronic systems, this behaviour describes the capacity of neurons to generate an output spike (action potential) once a threshold input has been reached.[46] Although we will not discuss the details of behavioural models here, the key functional importance of spiking is that it provides extremely low-power communication between neurons,[47] wherein the synaptic weight (the strength of the connection) determines the likelihood of an output being generated in response to a given input.[48] To explore the neuromorphic dynamics of $SiO_x$ at the nanoscale, we applied constant-current stresses at fixed locations using CAFM, to create filaments and push them into voltage instability without saturating the current detector and risking hard breakdown. The target current was set, then the voltage limit was increased at around 1 $Vs^{-1}$, with the applied voltage sampled at 200 Hz. Currents ranging from 10 pA to 10 nA were sampled at an instrumental sensitivity of 1 $nAV^{-1}$, and those ranging from 25 nA to 100 nA were sampled at a sensitivity of 100 $nA^{-1}$. As shown in **Figure 5**a, once the voltage limit is high enough, the current jumps above its target as a filament is formed. The applied voltage is then reduced to reduce the current to its setpoint. However, as is clear in Figure 5a, the voltage required to maintain the current setpoint is not stable, and we observe many spikes of varying magnitude.

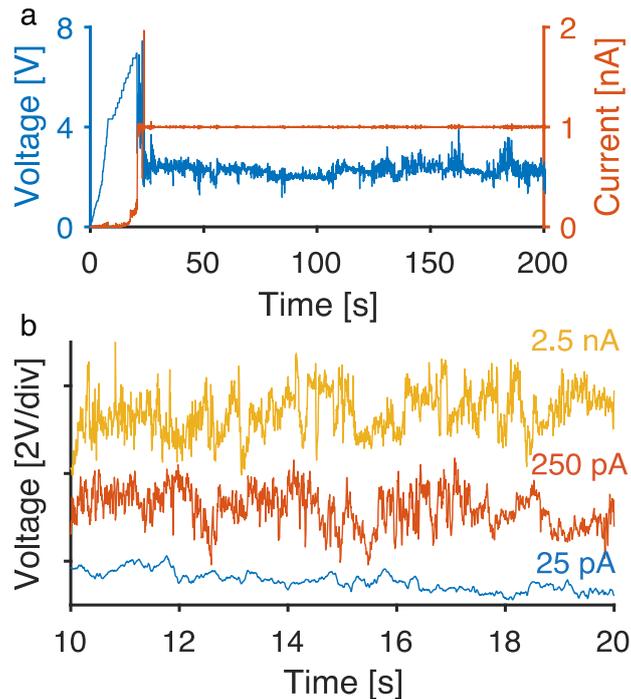

**Figure 5.** Spiking dynamics under constant current bias. No data processing filter has been applied. a) The voltage response at a setpoint of 1 nA. Once the target current has been reached, the voltage spikes consistently for hundreds of seconds. b) Spiking behaviour at different current biases over 10 seconds. As the current setpoint is increased, we observe higher magnitude, more frequent spikes. The difference between 25 pA and 250 pA is clear, although between 250 pA and 2.5 nA it is less distinct.

Figure 5b shows the spiking dynamics taken from 10 second windows at three constant current biases. At 25 pA, the trace does not show many, if any, spikes. Rather, the voltage feedback is effectively 'resting,' and maintaining the current setpoint without any significant changes to the filament. However, increasing the setpoint to 250 pA produces clear, frequent spikes. This is reasonable, as a higher current will increase the likelihood of disrupting the filament via Joule heating. Each time the filament begins to break, its conductivity will decrease, and so the voltage must increase in order to reform the filament and maintain the setpoint. Thus, a spike will occur. Higher currents will produce greater Joule heating, and so we would expect the filament to be broken more quickly, increasing the frequency of spikes. However, it is not clear from Figure 5b how this behaviour changes when the setpoint is increased from 250 pA to 2.5 nA.

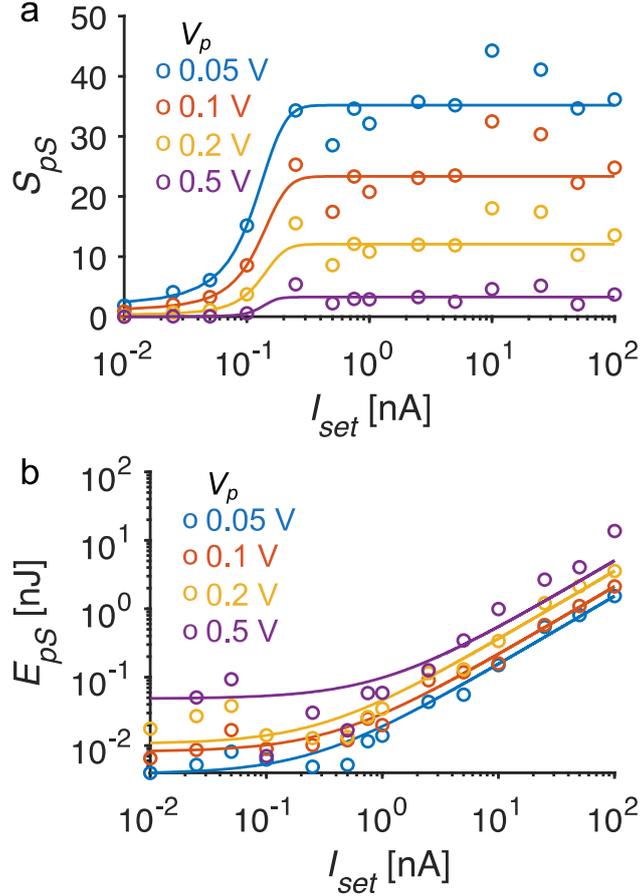

**Figure 6.** Characterization of neuromorphic behaviour in our $SiO_x$ for a range of prominences, $V_p$. a) Mean spikes per second, $S_{pS}$, as a function of current bias, $I_{set}$. At lower detection thresholds, $V_p$, we detect more frequent spikes. Solid lines are hyperbolic tangent fits to the data. Above 250 pA, $S_{pS}$ plateaus for all $V_p$. b) Energy per spike, $E_{pS}$, as a function of current bias, $I_{set}$. Solid lines are linear fits to the data. Below 1 nA, $E_{pS}$ is in the range of a few tens of pJ for $V_p$ up to 0.2 V.

To more quantitatively study the spiking dynamics in our $SiO_x$, in **Figure 6**a we plot the mean spikes per second, $S_{pS}$, defined as the total spikes counted in a measurement divided by the duration of counting in seconds, at a range of spike prominences, $V_p$. This corresponds to our detection threshold, i.e. we are determining whether or not to classify a given change in voltage as a spike (note that our CAFM produces a baseline noise of 0.4 mV when reading the sample bias at a fixed current setpoint, so our peak classification is statistically robust). Functionally, this represents the sensitivity of a connected component, such as another RRAM device, to input spikes. For all $V_p$, at low current bias, $S_{pS}$ is low, indicating that the voltage is in its resting state, with a reset unlikely to occur. As the current bias is increased, $S_{pS}$ also increases, as the filament is broken more frequently. This is a demonstration of a thresholding effect, wherein a low input current does not produce any output activation. Above 250 pA, a plateau is reached, with no further change as the current setpoint is increased by almost a further three orders of magnitude. We note that this plateaus in $S_{pS}$ is at around 35 per second, which is in good agreement with spike rates of 1 Hz to 100 Hz reported in the literature, as well as our previous observations.[24,49] As we increase $V_p$, $S_{pS}$ is lower at all values of $I_{set}$, though the shape of the trend remains consistent. This simply demonstrates that we record spikes more frequently if we have a lower detection threshold, as would be expected.

The data in Figure 6a are, in all cases, fitted well by a hyperbolic tangent, as in Equation 1:

$$S_{pS} = \frac{1}{2}S_{max}\{1 + tanh(p[I_{set} - I_0])\} \tag{1}$$

where $S_{max}$ is the maximum value of $S_{pS}$ for each $V_p$, $I_0$ is the threshold current to transition to a spiking state (i.e. $S_{pS} > \frac{1}{2} S_{max}$), and $p$ characterizes the sharpness of this transition. We have previously used a hyperbolic tangent to model the reset coefficient in our devices, and would therefore expect a similar shape for $S_{pS}$.[24] However, we note that a logistic sigmoid was also a good fit for the data in Figure 6a. Both of these functions are continuous and non-linear, and are commonly implemented in transformations between layers in convolutional neural networks.[50] Thus, functionally, $SiO_x$ appears able to produce a weighted output in response to inputs of varying magnitude. Phenomenologically, such logistic functions resemble the on/off activation and deactivation coefficients of the axonal ion pumps during generation of an action potential in a biological neuron.[49,51] While a full functional and phenomenological classification of the spiking behaviour of our devices with CAFM is beyond the scope of this work, it is very promising that we are able to observe these dynamics at the scale of individual filaments,[10,52,53] and at three orders of magnitude lower current than we have previously observed.[24] Crucially, these observations confirm that neuromorphic functionality in RRAM devices can be a highly localized phenomenon, and is further encouragement that this behaviour should persist as devices are scaled down.

It should be mentioned that we observed some topographical changes to the sample at the location of the probe contact point as a result of constant current stress measurements (Supplementary Figure S7). Up to 10 nA, these are a few nm in height, appearing similar to those observed following other measurements (Supplementary Figure S6), though they are not present following every applied stress. The appearance of the surrounding $SiO_x$ also does not change significantly after the stress has been applied. As such, we do not expect significant probe blunting or platinum deposition to have occurred at lower currents, suspecting instead that the $SiO_x$ and/or molybdenum have deformed. Conversely, above 10 nA the surface distortions are taller and the appearance of the surrounding material changes significantly, with the surface features appearing much wider and smoother. Therefore, we are unable to rule out probe blunting as a result of platinum deposition at higher currents. However, it is important to note that these values of $I_{set}$ are well above the plateau in $S_{pS}$, which occurs at around 250 pA, i.e. they are outside of the optimal range of device operation, so do not correspond to useful behaviour in the present context. This suggests that platinum deposition is not responsible for the functionality discussed above, specifically a weighted output in response to the input current, following a logistic shape.

Information in a spiking neural network is encoded by the timing and frequency of spikes.[49] As an event-based rather than continuous-output paradigm, this offers a very energy-efficient means of processing information, with both CMOS and RRAM devices demonstrating spikes of just a few fJ.[46,49,54,55] We have estimated the energy per spike, $E_{pS}$, in our CAFM measurements according to Equation 2:

$$E_{pS} = W_S \cdot V_S \cdot I_{set} \qquad (2)$$

where $W_S$ and $V_S$ are the full width at half maximum (which is log-normally distributed, so we have used the modal value) and height (which is normally distributed, so we have used the mean value) of spikes, respectively, at a current bias of $I_{set}$. For all $V_p$, $E_{pS}$ increases linearly with $I_{set}$. Up to around 1 nA current bias, we can see that $E_{pS}$ is up to a few tens of pJ, except for $V_p$ of 0.5 V, which is around 100 pJ as we are excluding the peaks of lower prominence and thus lower height. Therefore, for a device operating up to the $S_{pS}$ plateau observed in Figure 6a, we can expect the spike energy to be a few tens of pJ; increasing the current would not produce more spikes, though it would increase $E_{pS}$.

We expect that $E_{pS}$ would be reduced through optimization of the experimental setup. Using a CAFM probe as an electrode necessitates higher operational voltages, due to the high contact resistance that would not be present with a deposited top electrode,[56] and we have previously observed that less than 1 V may be used to switch devices with a top gold electrode and a 37 nm SiO$_x$ switching layer,[57] more than three times the thickness of the layer used in the present work. Furthermore, the voltage feedback of our microscope is slower than might be achieved with a semiconductor parameter analyser, with which we have observed switching processes to take place in tens of nanoseconds.[38] Therefore, we cautiously estimate, for a 100 nanosecond-wide spike at 1 V and 250 pA, an $E_{pS}$ of 25 aJ. This is a highly efficient event, representing the communication of information between devices, or changing in weight of a particular connection, with an energy consumption two orders of magnitude lower than that of conventional CMOS or other RRAM devices.[46,49,54,55] We are able to propose such a low value because the threshold for our $SiO_x$ to enter into a spiking state can be a very low current. If we consider an ideal, lossless, 1 cm × 1 cm RRAM array composed of 100 nm × 100 nm devices (i.e. around $25 \times 10^8$ devices), each firing at a rate of 10 Hz, an $E_{pS}$ of 25 aJ leads to a hypothetical power consumption of 0.6 μWcm$^{-2}$, more than four orders of magnitude lower than that of the human brain, which consumes around 10 mWcm$^{-2}$.[58] While this is hypothetical, as any real system will have a reduced efficiency due to line losses and device variability, for example, this is nevertheless a very exciting prospect for low power, high density neuromorphic computation.

## 4. Conclusions

In this work, we have demonstrated the use of CAFM as a tool for studying the plasticity and neuromorphic dynamics of $SiO_x$ at the nanoscale. The behaviour that we observe is in line with that of larger devices, although at a significantly lower energy of 25 aJ, which is very encouraging from the perspective of scaling and power consumption to produce significantly more efficient neuromorphic devices than those of conventional CMOS or other RRAM. We have additionally shown that not only do filamentary locations demonstrate conductivity enhancement, but that the surrounding background material also becomes more conductive under electrical stress. This core/shell behaviour is an important finding, supporting the conclusions of previous work and suggesting that this parallel nature of RRAM devices could be the source of rich dynamic behaviour that may be exploited in novel neuromorphic systems.

We have established that the instabilities in the behaviour of individual filaments are well-suited to neuromorphic functionality. The interplay of filament formation and rupture produces gradual plasticity under repeated electrical stressing, as well as voltage spiking under constant current stress. In the case of spiking, $SiO_x$ exhibits a current dependence that is fitted well by a continuous, logistic-type function, which suggests promising behaviour for neuromorphic computation. However, a full classification of the nanoscale neuromorphic behaviour of our $SiO_x$ is certainly necessary future work to fully understand these phenomena and further develop their functionalities. In particular, a thorough investigation of spike dynamics in nanoscale devices, including consideration of any background components is essential to fully understand their capacity for low-energy, high density neuromorphic applications.

## Acknowledgements

This work was supported by the Engineering and Physical Sciences Research Council, UK (grant no. EP/K01739X/1) Leverhulme Trust (grant no. RPG-2016-135) and The Worshipful Company of Scientific Instrument Makers.

# Supplementary Information

**Nanoscale plasticity and neuromorphic dynamics in silicon suboxide RRAM**

*Mark Buckwell\*, Wing H. Ng, Daniel J. Mannion, Stephen Hudziak, Adnan Mehonic, and Anthony J. Kenyon*

Here we present topography images associated with the current images shown in the manuscript. We did not observe significant damage to the probe or sample, other than when we applied high currents (tens of nA, Figure S7). In the context of voltage spiking, such currents are significantly higher than those required to reach the maximum value of $S_{pS}$ (250 pA). Thus, such large features correspond to the device being pushed beyond useful functionality. Otherwise, we observe a small (few nm in height) structural changes, such as small bumps (Figure S6 and Figure S7), and smoothing of the topography (Figure S1 and Figure S5). Topographical smoothing indicates either that the sample has deformed or that the probe has blunted. However, we do not expect the latter of these to have played a significant role in our measurements, as the appearance of the unstressed $SiO_x$ is generally consistent before and after electrical stress has been applied (Figure S3, Figure S6 and Figure S7), i.e. the probe condition is maintained. A possible cause of surface bumps is deposition of platinum from the probe onto the sample. However, we note that the appearance of structural changes on the $SiO_x$ surface is quite stochastic. In some cases, a few pA is sufficient to leave a residual feature (Figure S6), whereas in other cases, a current three orders of magnitude greater, 5 nA, causes no apparent change to the sample (Figure S7). Furthermore, these features can disappear over time (Figure S5 and Figure S6). We therefore expect that the observed surface changes indicate deformation of the $SiO_x$ and/or molybdenum as a result of electrical stress, as discussed extensively in the literature, or charging artefacts.

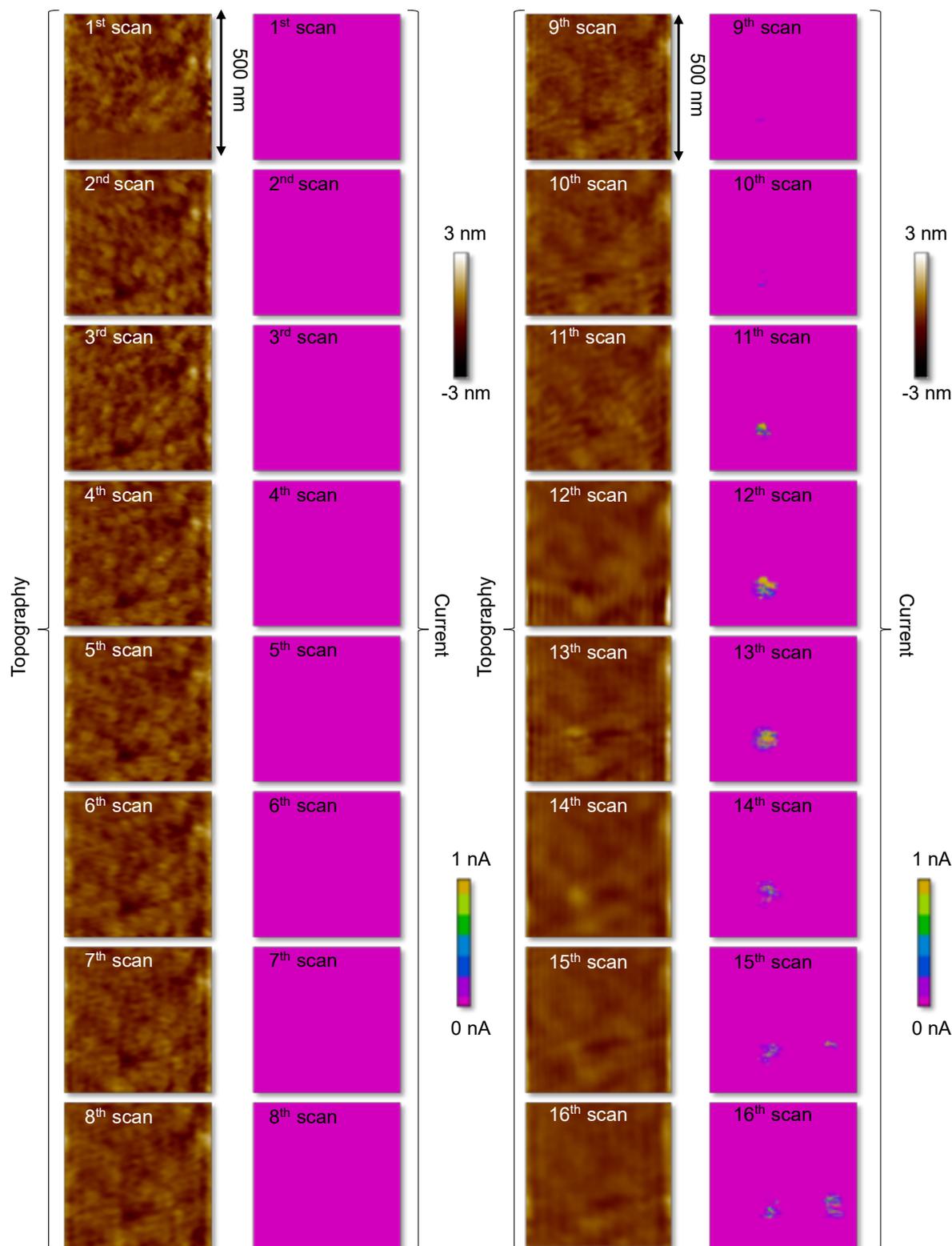

**Figure S1.** Full set of topography and current maps for the measurement presented in Figure 1 of the manuscript, when scanning at a constant bias of 7.7 V. The topography becomes gradually smoother, indicating either that the probe has become blunt or that the SiO$_x$ surface has deformed under the electrical stress, although the probe condition is maintained (Figure 3) so we do not expect significant blunting to have occurred. The current maps demonstrate the gradual potentiation and depression of the left-hand filament, and the eventual beginning of the potentiation of the right-hand filament.

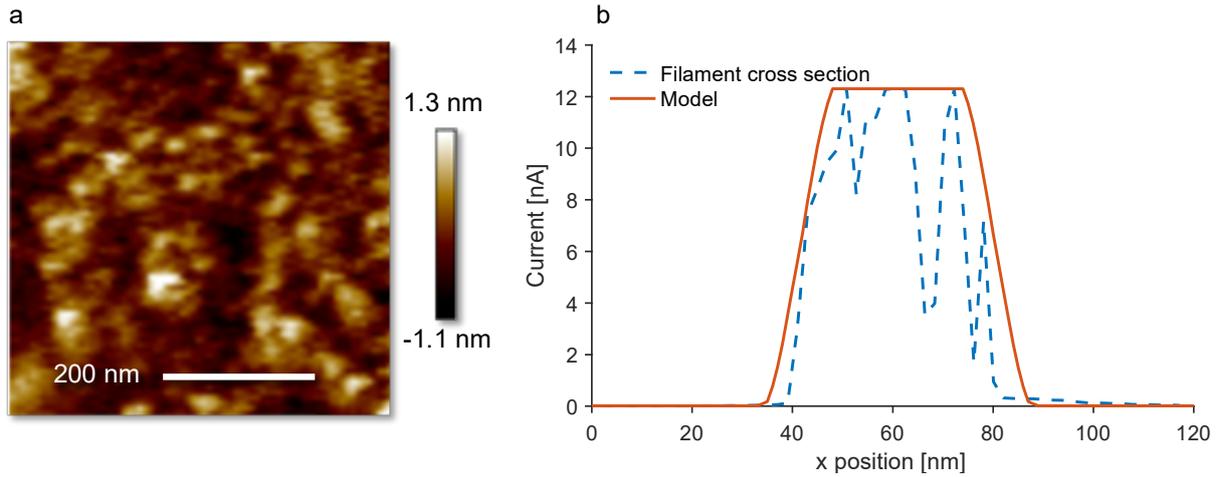

**Figure S2.** The methodology of estimating the filament width. a) Topography image of $SiO_x$ taken following the measurements shown in Figure 1 of the manuscript. This image was used to estimate the width of the probe to be 40 nm, using Nanoscope Analysis. b) A cross section was taken from the 13$^{th}$ scan (as shown in Figure S1), through the left-hand filament, at its widest point, so as to avoid underestimating the width. A 1D overlap addition was then performed between a probe 40 nm in length and of magnitude 1, and a conductive feature, modeled as a line of length $l$ and magnitude 1, to produce a convoluted trace. $l$ was then varied and found to best fit the cross-sectional data when set to 14 nm. The fitted data were then scaled to the maximum current measured in the filament. We stress that this is an estimate, to give a rough idea of how wide a feature might be if we assume a linear edge convolution. Notably, the real convolution would not be linear due to the geometry of the probe apex and the nonlinearity of the dependence of the current on the contact area. We might therefore expect the real filament to be narrower than 14 nm, or to be composed of multiple narrower filaments. Additionally, the current reached saturation, so the feature plateaus, which likely further convolutes its width. The image in a also serves as validation that the CAFM probe was still in good condition (i.e. not significantly blunted) following the measurements made in Figure 1 of the manuscript. This is because we are able to resolve features of a few tens of nm diameter, a similar size to the apex of a fresh the CAFM. However, we note that there may be a small tip artefact present, given the shape of the surface features of the $SiO_x$.

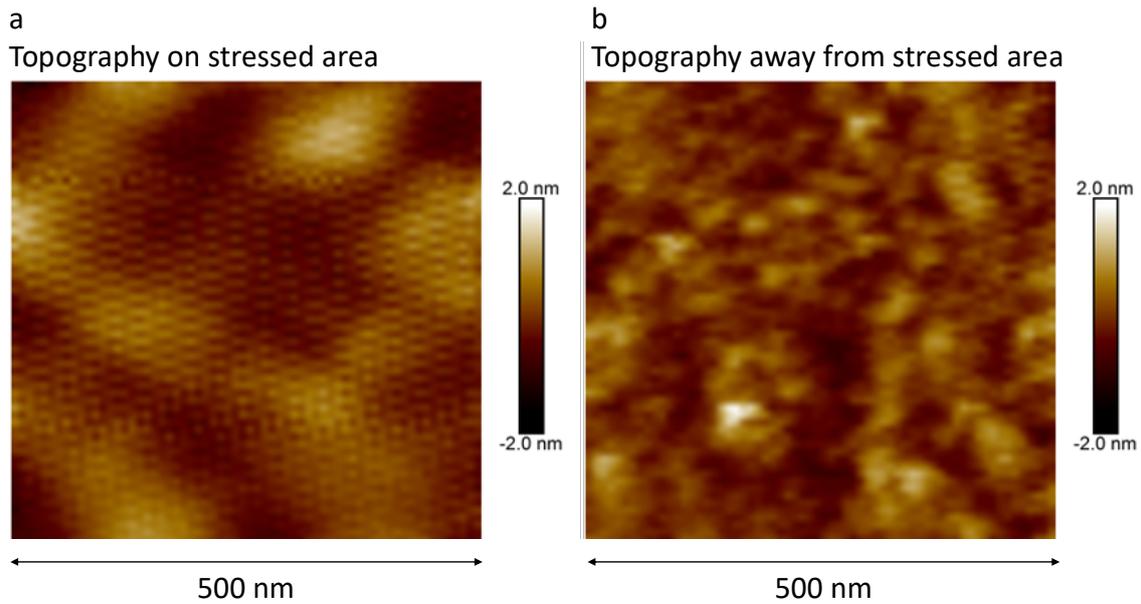

**Figure S3**. Topography scans demonstrating that the changing surface topography likely results from deformation of the $SiO_x$ and/or molybdenum, rather than changes to the probe, such as blunting. a) Topography of the area scanned in Figure 1, Figure 2 and Figure 3 of the manuscript, showing that the sample surface is rather distorted. b) Topography scan of an unstressed region of $SiO_x$, using the same probe as in a, immediately after the stressing was completed (i.e. after image a was taken). The surface appears similar to other pristine areas shown in the Supplementary Information, for example in Figure S4 and Figure S5. Thus, we do not expect the probe to have become significantly blunted during electrical stressing, and so do not expect that deposition of platinum has played a significant role in our observations, although we cannot rule this out. More likely, we suppose that the $SiO_x$ or molybdenum have deformed under electrical stress.

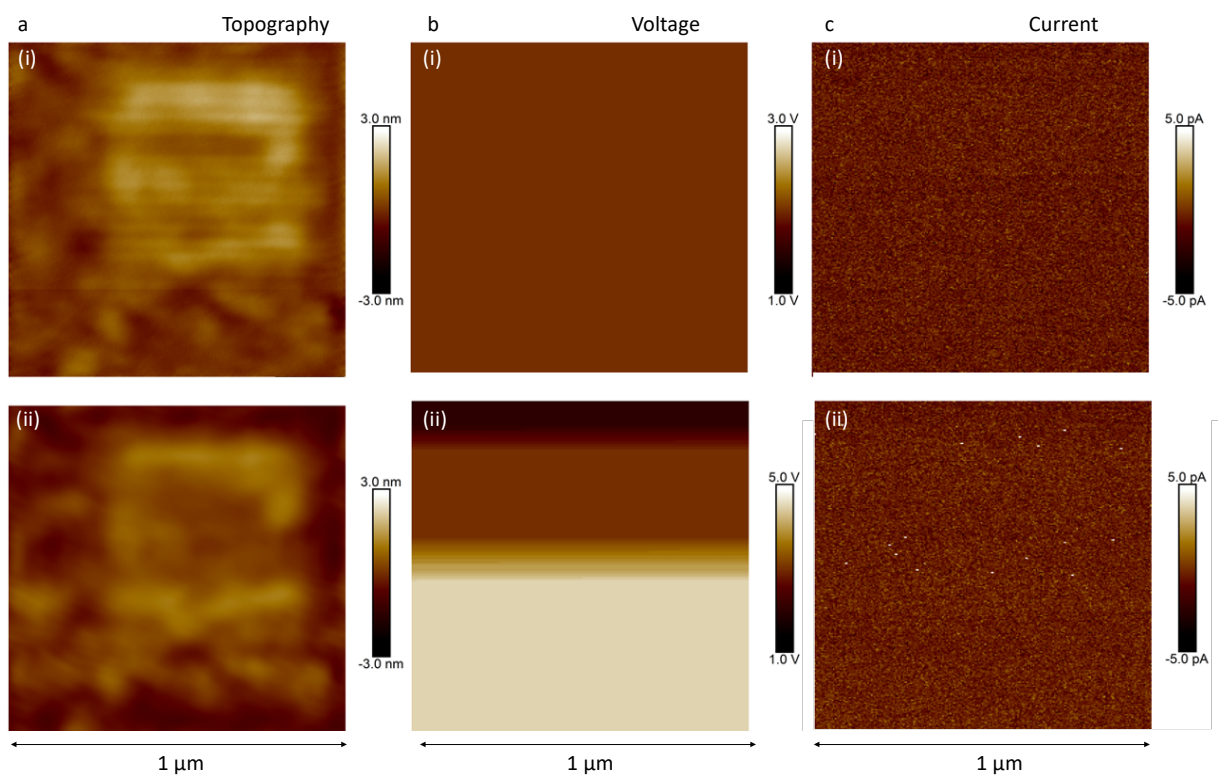

**Figure S4.** Two sets (i and ii) of topography (a), voltage (b) and current (c) scans taken on the region stressed in Figure 1, Figure 2 and Figure 3 of the manuscript, as well as Figure S5, demonstrating that voltages up to 5 V did not produce any measurable current. Note that there are some bright points in c (ii), corresponding to instrumental artefacts occurring when the applied voltage was changed.

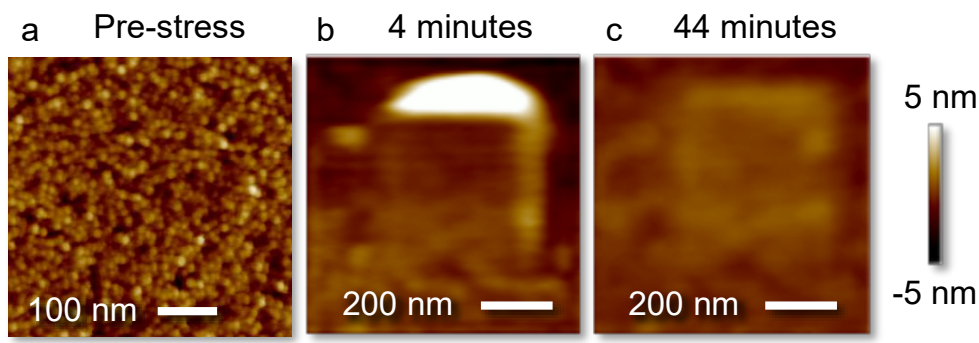

**Figure S5.** Topography of region scanned in Figure 1, Figure 2 and Figure 3 of the manuscript. a) Topography before any stress was applied, demonstrating the intrinsic structure of the SiO$_x$ and sharpness of the CAFM probe. b) A wider view of the same region, following 16 scans at 7.7 V, with the bias reduced to 5.5 V. It is clear that the either the surface has become smoother, or the probe has become blunter. Given the observations made in Figure S3, it is more likely that the surface has changed.. There is also a tall (around 5 nm) feature in the upper portion of the scan area. This might be a charging artefact, or surface deformation resulting from the repeated passage of the probe over the scan area. c) The same region as in b, following 40 minutes of scanning at 5.5 V. The topography does not appear to have changed significantly, although the raised feature in the upper portion of the scan area has disappeared.

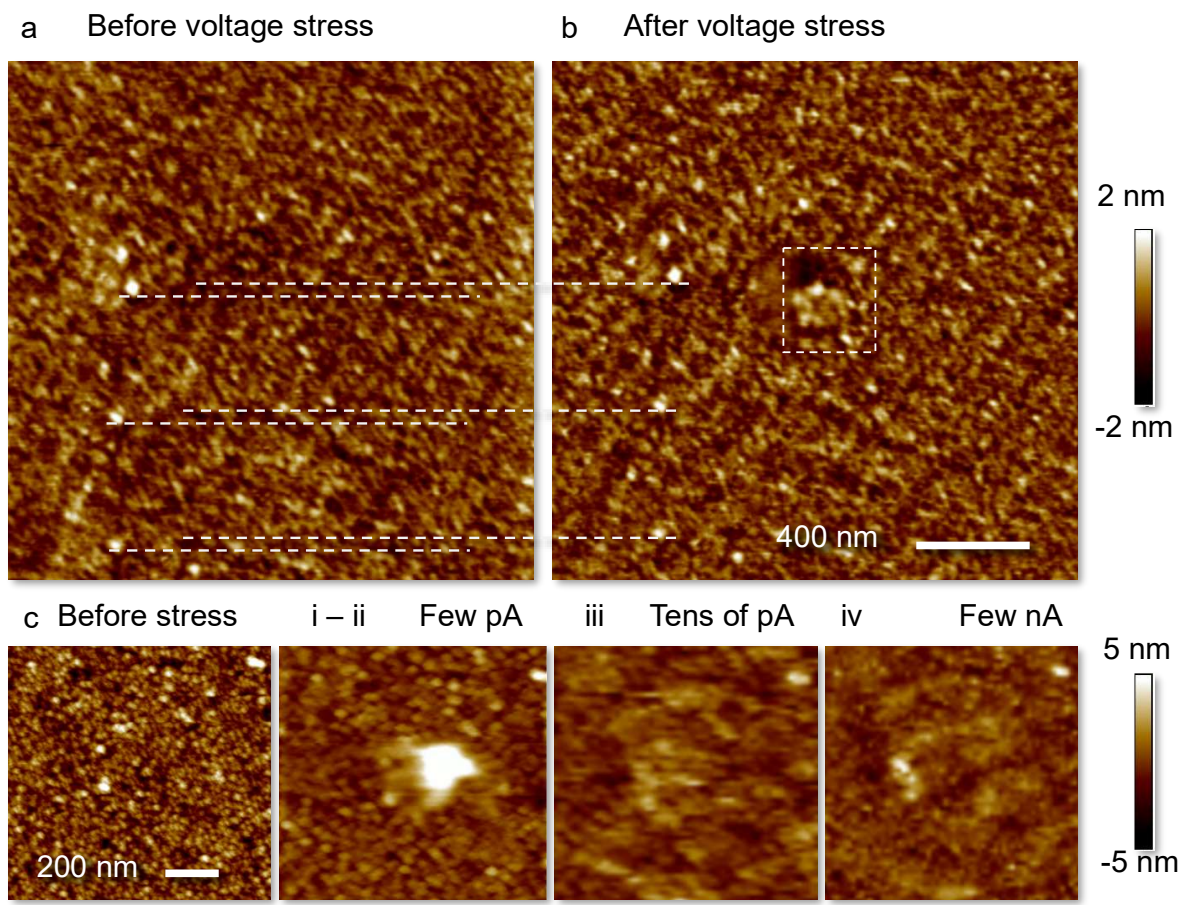

**Figure S6.** Topography of regions stressed in Figure 4 of the manuscript. a) and b) Topography before and after, respectively, the application of 4 V for around 1400 seconds, as shown in Figure 4a of the manuscript. We can see here that around 50 nm of drift has occurred during the scan, as indicated by the dashed white lines. This might have been a gradual process, due to the long duration of the measurement, in which case the time taken for the current to increase might be overestimated, as the contact location was constantly changing. Indeed, the current might only have begun to change more rapidly once a preferential filament formation location was reached. It is also possible that this occurred when the current sensitivity was changed at the end of the 1400 seconds, before imaging the stressed spot. Notably, when we changed the current sensitivity after the scan size was set to 0 nm, we observed jumps in the height and friction signals, indicating that the probe moved both vertically and laterally. Thus, the drift may have occurred before or after the electrical stress was applied, as the sensitivity was kept constant. There is also a slight surface deformation at the probe contact point, as indicated by the dashed white box, resulting from the electrical stress of 4 V and up to 12.3 nA. c) Topography images from the current maps shown in Figure 4b to d of the manuscript. The largest structural change is noted following a few pA of electrical stress. Subsequently, this feature disappears and there are no further large features present following a few nA of stress. The appearance of this feature on the $SiO_x$ surface during these measurements likely indicates either a transient charging artefact surface deformation that relaxes over the course of the measurement. We do not expect them to correlate with material deposition from the probe, as the diameter of features that we are able to resolve around the contact point before and after the measurements is quite consistent, at a few tens of nm (i.e. a similar size to the apex of the CAFM probe).

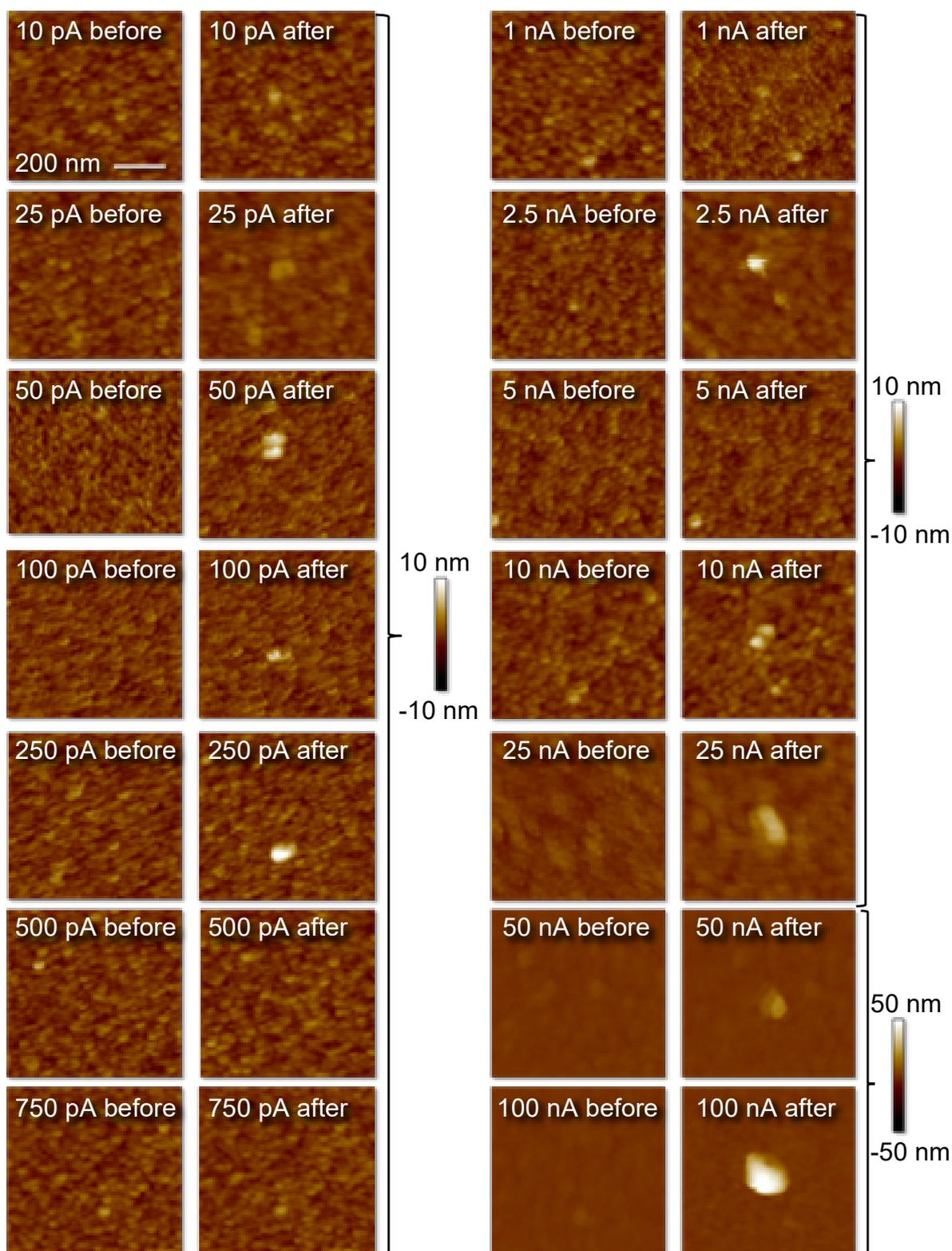

**Figure S7.** Topography images taken before and after the constant current measurements shown in Figure 5 and Figure 6 of the manuscript. We note that surface deformations appear as a result of current stress, although this seems to be a stochastic process, as not all stresses produce a deformation. Up to 25 nA, these features are at most a few nm in height. Above this, the features are tens of nm in height. However, as noted in Figure 6, $S_{pS}$ reaches a maximum from around 250 pA, and the topography data shown here indicates that significant structural deformation should only occur following the application of tens of nA.

31